# PV-VLM: A Multimodal Vision-Language Approach Incorporating Sky Images for Intra-Hour Photovoltaic Power Forecasting

Huapeng Lin, Miao Yu*

*College of Electrical Engineering, Zhejiang University, Hangzhou, 310027, China*

*Corresponding authors: zjuyumiao@zju.edu.cn (Miao Yu)*

**Abstract:** The rapid proliferation of solar energy has significantly expedited the integration of photovoltaic (PV) systems into contemporary power grids. Considering that the cloud dynamics frequently induce rapid fluctuations in solar irradiance, accurate intra-hour forecasting is critical for ensuring grid stability and facilitating effective energy management. To leverage complementary temporal, textual, and visual information, this paper has proposed PV-VLM, a multimodal forecasting framework that integrates temporal, textual, and visual information by three modules. The Time-Aware Module employed a PatchTST-inspired Transformer to capture both local and global dependencies in PV power time series. Meanwhile, the Prompt-Aware Module encodes textual prompts from historical statistics and dataset descriptors via a large language model. Additionally, the Vision-Aware Module utilizes a pretrained vision-language model to extract high-level semantic features from sky images, emphasizing cloud motion and irradiance fluctuations. The proposed PV-VLM is evaluated using data from a 30-kW rooftop array at Stanford University and through a transfer study on PV systems at the University of Wollongong in Australia. Comparative experiments reveal an average RMSE reduction of approximately 5% and a MAE improvement of nearly 6%, while the transfer study shows average RMSE and MAE reductions of about 7% and 9.5%, respectively. Overall, PV-VLM leverages complementary modalities to provide a robust solution for grid scheduling and energy market participation, enhancing the stability and reliability of PV integration.

## 1. Introduction

As the global economy continues to expand rapidly, fossil fuel consumption and climate change concerns are becoming increasingly pronounced [1]. Consequently, the advancement of renewable energy has emerged as a key strategy to tackle these challenges. Solar energy, characterized by its cleanliness, safety, and limitless availability, has been extensively utilized, with photovoltaic (PV) power generation is one of its principal applications. However, as PV power plants are integrated into the grid on a large scale, they continue to encounter issues related to output power instability. This instability is mainly driven by the natural variability of solar energy resources, with weather conditions exerting a particularly significant impact. The dynamic movement of clouds over PV panels results in rapid fluctuations in solar energy input, causing intermittent power generation and volatility, thereby posing serious challenges to the stable operation and economic scheduling of power systems. In this context, accurate forecasting of PV power generation is essential not only for improving system reliability [2] but also for enabling load aggregators to formulate well-informed strategies in the electricity market.

Numerous studies have demonstrated that extracting temporal features from sky images can significantly enhance forecasting performance. Approaches for extracting these features are typically divided into two categories: methods based on cloud motion derivation and deep learning-based techniques. Cloud motion derivation methods generate motion vectors by comparing consecutive sky images, thereby facilitating the extraction of critical parameters such as cloud speed, direction, and morphological changes. These parameters not only deepen our understanding of cloud evolution but also improve forecast accuracy. Common techniques for calculating cloud speed include particle image velocimetry (PIV), optical flow (OF) algorithms, and feature matching methods. Hu et al. [3] proposed an OF estimation method utilizing a recursive spatial pyramid convolutional neural network (CNN). By employing a multi-scale, recursive structure, the method effectively detects fine motion details in images. Although originally designed for video OF estimation, it also enables the extraction of cloud motion information, which forms a critical foundation for future applications in solar radiation and PV power forecasting. Liu et al. [4] focused on ultra-short-term solar radiation forecasting by leveraging sequential sky images to extract spatiotemporal features that capture cloud motion dynamics. The study employed the Scale-Invariant Feature Transform (SIFT) algorithm to construct sparse spatiotemporal feature descriptors. Zhen et al. [5] introduced an optimal weight combination method for cloud motion speed estimation, incorporating pattern classification and particle swarm optimization (PSO). Initially, they classified clouds in the sky using gray-level co-occurrence matrices and clustering techniques, followed by a weighted integration of block matching, OF, and feature matching to enhance cloud motion speed estimation accuracy. By accounting for the characteristics of different cloud types, this approach enables efficient cloud motion information extraction, thereby enhancing PV power forecasting accuracy. Karout et al. [6] incorporated direct normal irradiance (DNI) measurements with sky imaging data into their hybrid model for intra-hour DNI forecast. It is noteworthy that they utilized the Farneback OF algorithm to derive cloud motion features. The Farneback OF algorithm, a polynomial expansion-based OF estimation method, efficiently calculates pixel displacement in images, thereby accurate capturing cloud motion dynamics and supplying crucial spatiotemporal information to the model. Eşlik et al. [7] introduced a short-term solar radiation forecasting approach leveraging image processing and deep learning. Cloud motion was tracked using Shi-Tomasi feature detection and the Lucas-Kanade-OF method, while solar radiation forecasting was conducted using a long short-term memory (LSTM) network. Experiments confirmed that integrating cloud motion data into forecasting models reduces root mean square error (RMSE) in short-term solar radiation estimation, demonstrating its clear advantages.

Building on established methods for cloud motion estimation, recent studies have advanced predictive frameworks by integrating cutting-edge deep learning architectures and robust data processing techniques. These improvements enable the capture of more intricate spatiotemporal dynamics, which in turn enhance the accuracy of solar radiation and PV power forecasts. For instance, Xu et al. [8] introduced an innovative framework that combines a vision Transformer with a gated recurrent unit (GRU) encoder to capture high-dimensional spatiotemporal latent features from sky images. Their approach efficiently captures cloud motion dynamics over PV plants and integrates a multilayer perceptron module to enable one-step PV power prediction. This approach highlights the potential of Transformer architectures in enhancing ultra-short-term forecasting accuracy. In a separate study, Paulescu et al. [9] refined a hybrid forecasting model utilizing sky imaging technology to improve PV power estimation under clear-sky conditions. The model incorporates real-time physical adjustments to the predicted clear-sky output, considering the impact of cloud transmittance, thereby substantially enhancing intra-hour forecasting accuracy under rapidly varying sky conditions. Fu et al. [10] introduced a convolutional autoencoder (CAE)-based sky image forecasting model for minute-level PV power forecasting. Unlike traditional methods that rely on linear extrapolation, this model learns complex spatiotemporal patterns from historical sky images, enabling more accurate forecasts of future cloud distribution and PV power output. Furthermore, Nie et al. [11] examined the dataset imbalance problem in sky image-based forecasting. Given that clear-sky data tends to be predominant while cloudy-sky data is critical for accurate forecasting, they utilized resampling and data augmentation techniques within a CNN framework to address dataset imbalance. This approach not only enhanced PV power nowcasting accuracy but also significantly improved short-term forecasting several minutes into the future, emphasizing the critical role of comprehensive data preparation in solar energy prediction.

Current PV power forecasting methods typically process different modalities, such as sky images and power time series separately or through simple feature concatenation. Thus, they fail to capture the deep interactions among multimodal data, which are crucial for understanding cloud dynamics and real-time operating dynamics of the photovoltaic system. For example, sky images provide detailed insights into cloud motion and morphology, whereas PV power time series capture the real-time operational state of the system. Moreover, textual descriptions and physical model knowledge embed expert insights and prior information, both of which play a vital role in accurate power prediction. In intra-hour forecasting, the extended temporal scope of data and the increased complexity of weather variations and cloud motion make it challenging for single-modality or shallow fusion models to effectively capture the intricate relationships among key factors.

Vision-language models (VLMs) and large language models (LLMs) have recently shown remarkable promise in image analysis [12–14] and time-series forecasting [15–17], largely due to their ability to extract high-level semantic features and model complex temporal dependencies. Although VLMs excel at extracting rich visual information and LLMs effectively integrate domain expertise with nuanced semantic understanding, a systematic investigation into the potential of such cross-modal models for photovoltaic power forecasting is still lacking. Motivated by these developments, this study proposed a novel framework, PV-VLM, which integrates visual, temporal, and textual modalities through specialized processing modules and a tailored multimodal fusion strategy. By leveraging a VLM to capture cloud dynamics, a LLM to incorporate domain knowledge and a Transformer to extract local and global dependencies in PV power time series, the proposed approach significantly enhances the robustness and accuracy of intra-hour PV forecasting.

The contributions of this study are listed as follows.

(1) A novel framework is designed to integrate visual, textual, and time-series data for photovoltaic forecasting, ensuring a holistic representation of system behavior and capturing critical insights across modalities.

(2) A dedicated module employs a pretrained vision-language model to extract high-level semantic and spatial features from sky images, while a large language model encodes dynamic textual prompts from historical statistics and dataset descriptions to infuse rich domain knowledge.

(3) Leveraging a PatchTST-inspired Transformer to capture both local and global temporal dependencies, the framework introduces a cross-modal attention mechanism that adaptively fuses visual, textual, and temporal embeddings based on their relevance to PV power dynamics.

(4) Comprehensive comparative experiments and transfer studies across diverse climatic and geographical settings demonstrate significant reductions in RMSE and MAE, confirming the framework's enhanced robustness and practical value in real-world PV forecasting.

**2. Method**

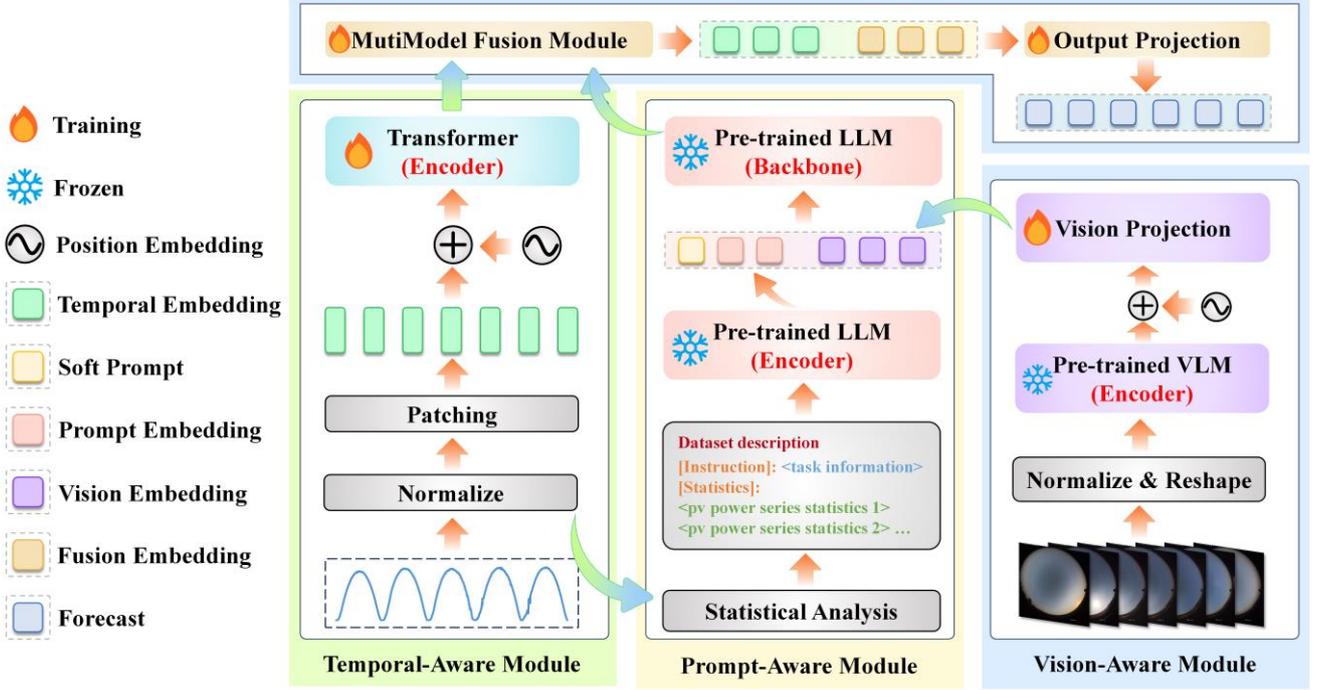

**Fig. 1 Framework of the proposed multimodal pipeline for PV power forecasting**

As shown in Fig. 1, we present a multimodal pipeline for PV power forecasting by integrating sky image features, textual prompts, and time-series embeddings in a unified representation. Sky images are first passed through a VLM to obtain high-level visual embeddings, which are then refined via a vision projection layer for improved alignment. Simultaneously, statistical analysis of PV power data is performed to dynamically construct textual prompts, including learnable soft prompts, which are encoded by a LLM to generate textual embeddings. These refined visual and textual features are concatenated and input to the LLM backbone to produce an integrated multimodal embedding. Meanwhile, the historical PV power sequence is segmented into several patches and processed by a PatchTST-like [18] transformer encoder, capturing both local fluctuations and global dependencies in the time series. Finally, within the multimodal fusion module, the integrated LLM output is fused with the time-series embedding through a cross-modal attention mechanism, followed by a skip connection and layer normalization. The fused embedding is passed through an output projection layer to generate robust and accurate PV power forecasts. This design harnesses the complementary strengths of sky-image cues, domain-specific textual insights, and temporal patterns to enhance forecasting performance under diverse atmospheric conditions.

### 2.1. Vision-Aware Module (VAM)

In PV system power forecasting, sky images offer valuable visual information for inferring trends in solar irradiance changes. The VAM utilizes a pre-trained VLM to extract high-level features from sky images and map them to a shared feature space aligned with textual embeddings using the following operations:

**(1) Bilinear interpolation.** The bilinear interpolation is applied to resize the input image to the desired resolution $(H, W)$. For a target pixel $(x, y)$ in the resized image, its interpolated intensity value $I'(x, y)$ from the original image $I$ can be computed in a simplified from as

$$I'(x, y) = \sum_{i=1}^{2}\sum_{j=1}^{2} I(x_i, y_j) \cdot w_{ij} \tag{2-1}$$

where $(x, y)$ are the coordinates of the four nearest neighbors, $w_{ij}$ are the bilinear interpolation weights determined by the relative distances from the target pixel.

**(2) Normalization.** After interpolation, the image is normalized to ensure consistency with the distribution expected by the pre-trained visual encoder. First, we scale the pixel values to the range [0,1] by dividing by 255 as

$$I_{norm}(x, y, c) = \frac{I'(x, y, c)}{255} \tag{2-2}$$

where $c$ represents the color channel.

**(3) Standardization.** Standardization. Each color channel is further standardized using the pre-trained encoder's mean $\mu_c$ and standard deviation $\sigma_c$ for each channel

$$I_{std}(x, y, c) = \frac{I_{norm}(x, y, c) - \mu_c}{\sigma_c} \tag{2-3}$$

**(4) Visual encoding.** After applying bilinear interpolation and standardization, the image $I_{std}$ is first fed into a VLM, such as CLIP [19], BLIP-2 [20], or SigLIP-2 [21], to extract high-level visual semantic and spatial information. The VLM processes the input image through its vision encoder, which consists of convolutional layers or transformer-based feature extractors, to extract high-level visual features. These features capture semantic information about the scene, including cues related to cloud formations and their movement, which can support further analysis for PV power forecasting. The output of the VLM, denoted as $F_{VLM}$, contains a rich set of visual features that preserve both local textures and global scene information.

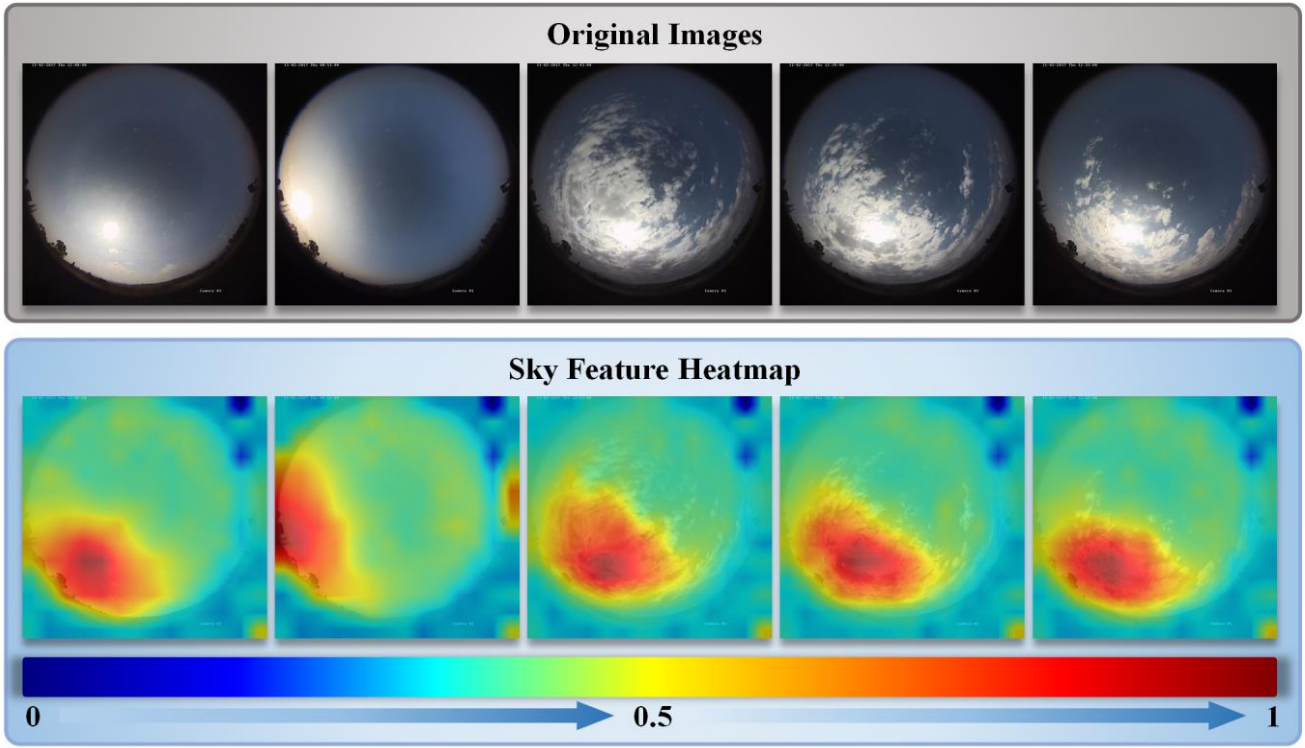

**Fig. 2 Heatmap of high-level semantic features extracted by SigLIP-2 for a sky image.**

To further demonstrate the effectiveness of the pretrained visual encoder, Fig. 2 presents a heatmap visualization of semantic feature activations extracted by SigLIP-2. Notably, without requiring any additional fine-tuning, the pretrained model is able to identify visually salient regions such as cloud edges, dense formations, and irradiance gradients that are highly relevant for PV power fluctuations. These activation patterns correspond to areas with rapid changes in solar input, highlighting the model's inherent capacity to capture predictive visual cues. This confirms that pretrained vision-language models, even when frozen, can effectively support intra-hour PV forecasting by contributing spatial context critical to understanding irradiance dynamics.

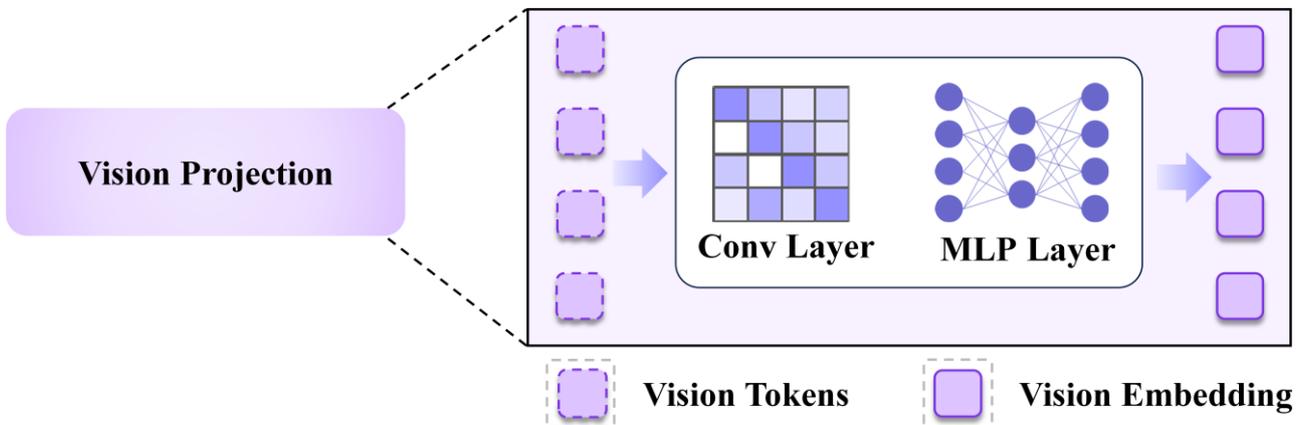

**Fig. 3 Architecture of vision projection in VAM**

**(5) Vision Projection.** Following VLM feature extraction, the output features $F_{VLM}$ are passed into a vision projection layer, as shown in Fig. 3, which performs dimensionality reduction and aligns the extracted visual features via a convolutional layer followed by a multi-layer perceptron (MLP). In addition, a sinusoidal positional embedding is combined with $F_{VLM}$ to inject spatial information into the visual features [22]. This transformation can be expressed as:

$$F_{conv} = \text{ConvLayer}(F_{VLM}) \tag{2-4}$$

$$E_{img} = \text{MLP}(F_{conv}) \tag{2-5}$$

where $\text{ConvLayer}(\cdot)$ represents the convolutional operation applied to the VLM output, which refines the spatial structure of extracted features to improve local feature consistency and alignment. $\text{MLP}(\cdot)$ denotes the multi-layer perceptron responsible for projecting the transformed visual embeddings into a multimodal hidden space that aligns with the language model and time-series representations.

The vision projection module ensures that the extracted image features are not only refined but also structured for seamless fusion with textual embeddings. By extracting high-level semantic features from the VLM and subsequently aligning them with textual data through convolution and MLP-based projection, the model effectively integrates sky image information into the PV power forecasting pipeline, enhancing predictive accuracy and robustness in scenarios with dynamic cloud movement.

## 2.2. Prompt-Aware Module (PAM)

> **[Input power description]:** This pv power dataset collected over the past 3 years. The dataset represents clear daily periodicity …
> **[Input images description]:** these are sky images of size 64 taken by a ground-based fish-eye camera …
> ----
> Below is the information about the input time series:
> **[Instruction]:** forecast the next \<Horizon\> steps given the previous \<Input Size\> steps information;
> **[Statistics]:** The input has a minimum of \<min_val\>, a maximum of \<max_val\>, and a median of \<median_val\>. The overall trend is \<upward or downward\>. The top five lags are \<lag_val\>.

**Fig. 4 Prompt example.** \< \> and \< \> are task-specific configurations and calculated input statistics

Inspired by the prompt-as-prefix approach in Time-LLM [23], the PAM module generates contextual textual feature specifically for PV power forecasting, using dynamically derived information from PV power time series data. A prompt example is shown in Fig. 4. These dynamically generated prompts focus on key statistical attributes, including:

(1) Range: Minimum and maximum values of PV power.

(2) Central Tendency: Median PV power output.

(3) Trend Analysis: Directional tendencies derived from first-order differences.

(4) Task Description: Defined forecasting horizons and historical data lengths.

(5) Dataset Description: Detailed descriptions specific to PV systems, including information on sky imaging camera specifications.

In addition, we introduce a soft prompt as a complementary mechanism. A soft prompt is a set of learnable prompt embeddings that are not fixed text but rather parameterized vectors optimized during training [24]. This mechanism automatically captures latent semantic information within the data and provides LLM with additional contextual cues. The soft prompt enhances the integration of dynamically generated textual information with visual features, ultimately improving the accuracy and robustness of PV power forecasting.

By incorporating this dynamically tailored dataset context, the LLM gains an improved understanding of the unique temporal patterns and characteristics inherent to PV power datasets. The structured statistical attributes and domain-specific task definitions serve as critical guidance, ensuring that the LLM correctly maps embeddings for accurate forecasting.

**(1) Textual context encoding.** The dynamically generated textual contexts are tokenized and transformed into embeddings via the LLM text encoder, which can be instantiated as GPT-2 [25], BERT [26], Qwen2.5 [27], or similar large-scale language models. Given a prompt $T$ dynamically constructed from the statistical attributes, the LLM encodes the text as:

$$E_{text} = \text{LLM}_{encoder}(T) \tag{2-6}$$

**(2) Text–Vision Embedding Concatenation.** To effectively integrate both textual and visual features, these embeddings are concatenated before being fed into the LLM:

$$E_{mm} = \text{Concat}(E_{text}, E_{img}) \tag{2-7}$$

**(3) Preliminary Multimodal Representation.** The fused embedding $E_{mm}$ is then processed by the LLM's backbone:

$$E_{LLM} = \text{LLM}(E_{mm}) \tag{2-8}$$

This multimodal feature $E_{LLM}$ serves as the final feature, fusing both statistical textual context and sky image features. By integrating structured prompts with vision-derived spatial insights, the module achieves a comprehensive understanding of PV power dynamics, improving PV power forecasting accuracy and robustness.

### 2.3. Temporal-Aware Module (TAM)

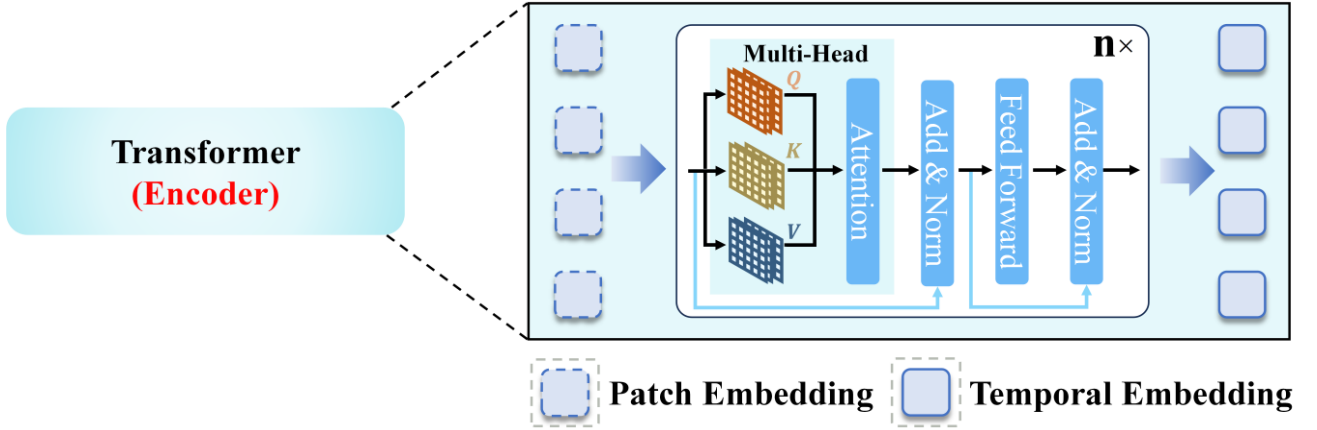

**Fig. 5 Architecture of Transformer encoder**

As shown in Fig. 5, the TAM is responsible for extracting informative embeddings from PV power time series data. To effectively model both short-term fluctuations and long-term dependencies, TAM adopts a structure inspired by PatchTST, where time series data is segmented into several patches and processed using self-attention mechanisms. Unlike traditional recurrent or convolutional approaches, this patch-based modeling enhances the ability to capture both local patterns and global dependencies.

**(1) Patch segmentation.** Given a PV power time series $P$ of length $L$, the sequence is first partitioned into overlapping patches of size $L_p$ with a stride of $S$. This transformation reshapes the input into a sequence of patches:

$$X_{patch} = \text{Reshape}(P_t) \in \mathbb{R}^{B \times N_{patch} \times L_p} \tag{2-9}$$

where $B$ is the batch size, $N_{patch} = \frac{L-L_p}{S} + 1$ is the number of extracted patches, and $L_p$ represents the patch length.

**(2) Patch-level projection.** Each patch is then linearly projected into a latent feature space of dimension $d_{model}$:

$$F_{ts} = W_{ts} X_{patch} + b_{ts} \tag{2-10}$$

where $W_{ts} \in R^{L_p \times d_{model}}$ and $b_{ts} \in R^{d_{model}}$ are learnable parameters.

**(3) Temporal encoding via Transformer.** The patches are subsequently processed by a multi-layer transformer encoder:

$$E_{temporal} = \text{TransformerEncoder}(F_{ts}) \tag{2-11}$$

where the self-attention mechanism dynamically weights past observations, enabling the model to focus on the most relevant historical patterns that influence PV power fluctuations. The transformer encoder captures both short-term variations and long-term dependencies, making it well-suited for time series forecasting in complex environmental conditions. Unlike standard PatchTST, TAM does not include a separate head module; instead, the extracted temporal embeddings are directly integrated into the multimodal fusion process.

This approach allows the model to seamlessly incorporate temporal dependencies into a unified forecasting framework, enhancing its robustness in PV power prediction across diverse atmospheric conditions.

### 2.4. Multimodel Fusion Module (MMFM)

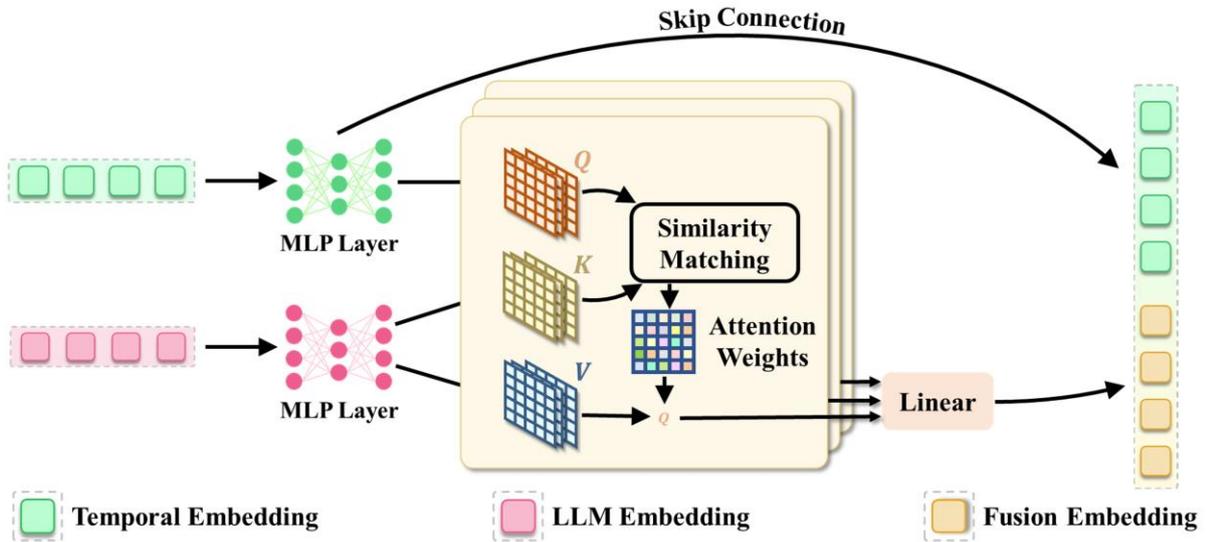

**Fig. 6 Architecture of MMFM**

The multimodal fusion pipeline integrates visual, textual and temporal embeddings leveraging their complementary strengths to enhance PV power forecasting, as shown in Fig. 6.

**(1) Modality projection into a shared latent space.** To ensure compatibility, both embeddings are projected into a shared latent space of dimension $d_{model}$ using learnable transformations

$$E'_{LLM} = W_{LLM} E_{LLM} + b_{LLM} \tag{2-12}$$

$$E'_{temporal} = W_{temporal} E_{temporal} + b_{temporal} \tag{2-13}$$

where $W_{LLM}, W_{temporal}$ are learnable weight matrices and $b_{LLM}, b_{temporal}$ are bias terms that align the embeddings in a unified latent space.

**(2) Cross-modal attention** To facilitate multimodal interaction, as show in Fig. 6, a cross-modal attention (CMA) mechanism is applied, where the $E'_{temporal}$ serves as the query while the multimodal representation from the $E'_{LLM}$ acts as the key and value, compute for each head $h$ (where h = 1, ..., H, with $H$ representing the total number of attention heads):

$$Q_h = E'_{temporal} W_h^Q, \quad K_h = E'_{LLM} W_h^K, \quad V_h = E'_{LLM} W_h^V \tag{2-14}$$

where $W_h^Q, W_h^K, W_h^V$ are learned projection matrices and $d_k$ is the dimensionality per head. The attention output for head $h$ is then

$$attn_h = \text{Softmax}(\frac{Q_h K_h^\top}{\sqrt{d_k}}) V_h \tag{2-15}$$

**(3) Multi-head aggregation.** The multi-head CMA output is obtained by concatenating the outputs from all heads

$$\text{CMA}(E'_{temporal}, E'_{LLM}) = \text{Concat}(attn_1, \ldots, attn_h) \tag{2-16}$$

**(4) Residual Skip Connection with Layer Normalization** To enhance feature retention and avoid information loss, a residual skip connection [28] is incorporated, followed by Layer Normalization to stabilize training and maintain numerical stability

$$E_{fusion} = \text{LayerNorm}(E'_{temporal} + \text{CMA}(E'_{temporal}, E'_{LLM})) \tag{2-17}$$

**(5) Output projection.** The final fused representation is then processed by an output projection layer, which comprises a flattening operation followed by a learnable full connected layer. This operation is formulated as

$$Forecast = \text{OutputProjection}(E_{fusion}) \tag{2-18}$$

This design ensures that the fusion module effectively integrates long-term dependencies from historical power data with rich semantic and spatial knowledge from the multimodal LLM backbone, thereby improving the robustness and accuracy of PV power forecasting.

## 3. Experimental Results

### 3.1. Dataset and metrics

In this study, two globally collected datasets from distinct locations with vastly different weather conditions are utilized to explore transfer learning capability. These datasets are:

**Dataset A:** This dataset, curated for deep-learning-based solar forecasting research at Stanford University in California, USA [29], includes three years of processed, down-sampled sky images with a resolution of 64 × 64 captured by a ground-based fish-eye camera. It also contains power output measurements from a 30-kW rooftop PV array located approximately 125 meters from the camera. Both data sources are logged at a 1-minute frequency.

**Dataset B:** This dataset, collected at the University of Wollongong in Wollongong, New South Wales, Australia [30], features sky images with a resolution of 1024 × 768 captured on the main campus between 8:00 a.m. and 4:45 p.m., along with corresponding PV power output measurements. All data streams are recorded at a 1-minute interval.

According to the Köppen-Geiger climate classification [31], Dataset A was collected from the region encompassing Stanford University, which exhibits a warm-summer Mediterranean climate, while Dataset B originates from a humid subtropical zone. Following extensive preprocessing and quality filtering, Dataset A comprises approximately 349,000 samples, whereas Dataset B contains around 6,000 samples. To ensure consistency and facilitate direct comparisons, only the 2019 data were retained in Dataset A, matching Dataset B's temporal coverage. Moreover, the original 1-minute data were resampled to a 2-minute interval, reducing training time while preserving the operational sampling granularity in PV power forecasting. Table 1 summarizes the overview of Datasets A and B after processing.

Table 1 Overview of Dataset A and Dataset B after processing.

| Dataset | A | B |
| --- | --- | --- |
| Location | Stanford University, California, USA | University of Wollongong, Australia |
| PV plants | 1 | 2 |
| Samples after preprocess | 62910 | 3000 |
| Climate | warm-summer Mediterranean | humid subtropical |
| Image size | 64x64 | 1024x768 |

The performance of the proposed framework is assessed using RMSE, Mean Absolute Error (MAE) and the Coefficient of Determination ($R^2$) as evaluation metrics. RMSE quantifies the average magnitude of errors in the same unit as the target variable, making it interpretable for practical applications. MAE measures the mean absolute deviation between predictions and actual values, providing an intuitive understanding of the model's accuracy without emphasizing large errors. $R^2$, on the other hand, represents the proportion of the variance in the dependent variable that is predictable from the independent variables, offering a measure of how well unseen samples are likely to be predicted by the model. The mathematical definitions of these metrics are given as follows:

$$RMSE = \sqrt{\frac{1}{N}\sum_{i=1}^{N}(\hat{y}_i - y_i)^2} \qquad (3\text{-}1)$$

$$MAE = \frac{1}{N}\sum_{i=1}^{N}|\hat{y}_i - y_i| \qquad (3\text{-}2)$$

$$R^2 = 1 - \frac{\sum_{i=1}^{n}(y_i - \hat{y}_i)^2}{\sum_{i=1}^{n}(y_i - \overline{y})^2} \tag{3-3}$$

where $y_i$, $\hat{y}_i$ and $\overline{y}$ represent the actual value, predicted value and mean of the actual values, respectively, and $N$ denotes the total number of samples. These metrics together offer a comprehensive assessment of the models' forecasting performance.

## 3.2. Baseline frameworks and experiment set-ups

The objectivity and reliability of the results were ensured by conducting three experiments and averaging their outcomes. Four recently proposed deep learning-based frameworks were employed as baselines: Multiple image convolutional long short-term memory fusion network (MICNN-L) [32], Bi-level spatial-temporal network (BILST) [33], Convolutional LSTM (ConvLSTM) [34], and Stanford University Neural Network for Solar Electricity Trend (SUNSET) [35]. Specifically, MICNN-L employs convolutional LSTM layers to fuse multi-temporal images, capturing subtle spatial details and dynamic changes over time that are crucial for solar power forecasting. BILST leverages a dual-level architecture where one level extracts global context while the other focuses on local nuances, enabling a more precise modeling of intricate spatiotemporal interactions. ConvLSTM integrates convolution operations into LSTM cells to maintain spatial integrity and sequential dynamics simultaneously, offering a straightforward yet effective approach. SUNSET is specifically tailored for solar electricity trend prediction by incorporating customized preprocessing and network components that address both periodic behaviors and sudden fluctuations inherent in solar data. Due to differences in datasets and limited computational resources, the baseline models were re-implemented strictly according to the methodological descriptions in their original publications, without additional dataset-specific adaptations or extensive hyperparameter tuning.

Our proposed framework integrates SigLIP-2 as the VLM to extract high-level semantic and spatial representations from sky images, ensuring that cloud patterns and irradiance variability are effectively encoded. For textual understanding, we employ GPT-2 as the LLM to process statistical prompts, capturing essential PV trends and generating structured embeddings that enhance multimodal fusion. The combination of SigLIP-2 and GPT-2 allows the model to leverage both spatial-temporal dependencies and domain-specific textual insights, contributing to improved PV power forecasting accuracy.

All these frameworks were developed on PyTorch 2.6.0 platform with Python 3.11.10. The Adam optimizer was chosen for parameter optimization and the mean square error (MSE) served as the loss function. A cross-validation strategy was applied, with 70% of the dataset allocated for training, 10% for validation, and 20% for testing. Moreover, all experiments were conducted on a cluster with 2048 GB of RAM, Intel Xeon 8468 CPU and NVIDIA H800 graphic card.

To ensure fair comparison, all models were trained using Early Stopping [36] to prevent overfitting, where training is halted if the validation loss does not improve for a predefined number of epochs, known as the patience value. In our experiments, the patience was set to 5. Additionally, a learning rate scheduler was applied to improve convergence by periodically reducing the learning rate according to the following update rule

$$\eta_t = \eta_0 \cdot \gamma^{\frac{t}{s}} \tag{3-4}$$

where $\eta_t$ is the learning rate at epoch $t$, $\eta_0$ is the initial learning rate, $\gamma$ is the decay factor, and $s$ is the step size.

For the proposed model, training was further optimized using the BFloat16 precision format to enhance computational efficiency and reduce memory usage. The main training hyperparameters are summarized in Table 2.

Table 2 Main training hyperparameters used in all experiments.

| Main hyperparameters | Value |
|---|---|
| Batch size | 32 |
| Epochs | 50 |
| Patience | 5 |
| $\gamma$ | 0.1 |
| $s$ | 3 |
| $\eta_0$ | 0.001 |

### 3.3. Comparison of different horizons

We focus on ultra-short-term forecasts of 10, 20, and 30 steps, forecasting PV power up to 20, 40, and 60 minutes ahead. The input lengths are twice the horizons, providing sufficient historical context to distinguish ultra-short-term fluctuations from underlying trends. This extended observation window helps the model capture temporal dependencies more effectively, leading to more stable and accurate forecasts while improving its adaptability to rapid irradiance changes.

Table 3 Comparisons between the proposed and baseline frameworks for different horizons. Red: the best, Blue: the second best.

| Methods | Proposed | | | MICNN-L | | | BILIST | | | ConvLSTM | | | SUNSET | | |
|---|---|---|---|---|---|---|---|---|---|---|---|---|---|---|---|
| Horizon(min) | 20 | 40 | 60 | 20 | 40 | 60 | 20 | 40 | 60 | 20 | 40 | 60 | 20 | 40 | 60 |
| RMSE(kW) | 1.604 | 1.767 | 1.925 | 1.666 | 1.918 | 2.044 | 1.720 | 1.897 | 2.043 | 1.738 | 1.879 | 2.094 | 5.893 | 5.311 | 4.699 |
| MAE(kW) | 0.698 | 0.857 | 0.947 | 0.703 | 0.931 | 1.096 | 0.950 | 1.194 | 1.195 | 0.887 | 0.876 | 1.152 | 4.928 | 4.504 | 3.970 |
| $R^2$(/) | 0.948 | 0.937 | 0.925 | 0.944 | 0.926 | 0.916 | 0.940 | 0.928 | 0.916 | 0.939 | 0.929 | 0.912 | 0.301 | 0.432 | 0.555 |

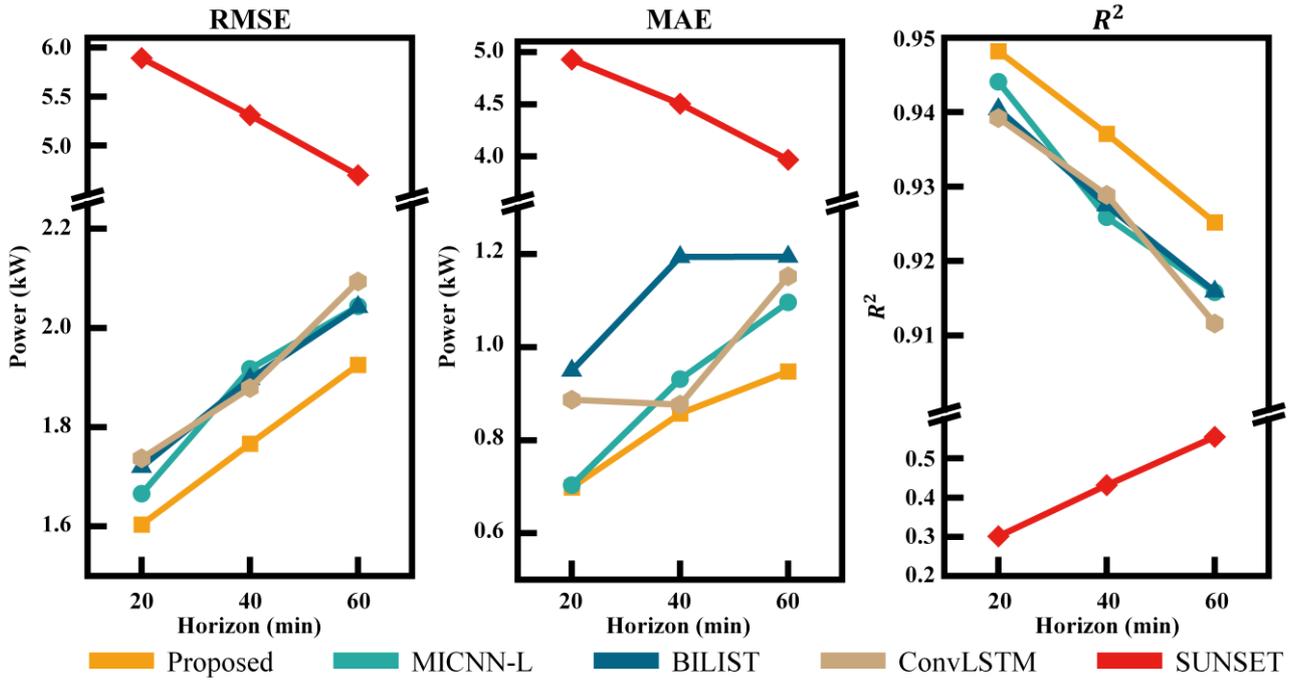

**Fig. 7 Performance metrics of the proposed method and baseline models across different forecast horizons.**

As shown in Table 3 and Fig. 7, on the 20-minute horizon, the proposed method achieves a 3.72% reduction in RMSE, a 0.71% reduction in MAE, and a 0.42% improvement in R² compared to the best competing model. For the 40-minute forecast, the improvements are even more pronounced, with RMSE reduced by 5.96%, MAE by 2.17%, and R² increased by 0.86%. On the 60-minute horizon, the proposed approach further outperforms the best competitor by reducing RMSE by 5.72% and MAE by 13.5%, while boosting R² by approximately 1%. Overall, these results demonstrate that the proposed method consistently delivers lower forecasting errors and enhanced accuracy across various horizons, effectively capturing the complex dynamic characteristics of PV power time series.

As illustrated in Fig. 8, forecast results for three selected days (representing sunny, cloudy, and overcast weather conditions) are presented. On sunny days, all models display predictions that closely follow the actual power output. Under cloudy and overcast conditions, the inherent weather uncertainty leads to larger deviations among the models. Nevertheless, the proposed method (highlighted in orange) consistently achieves better alignment with the actual power curves, both during peak power periods and low output intervals. This finding reinforces the statistical results shown in Table 3 and demonstrates the model's strong adaptability and generalization capability under varying weather conditions.

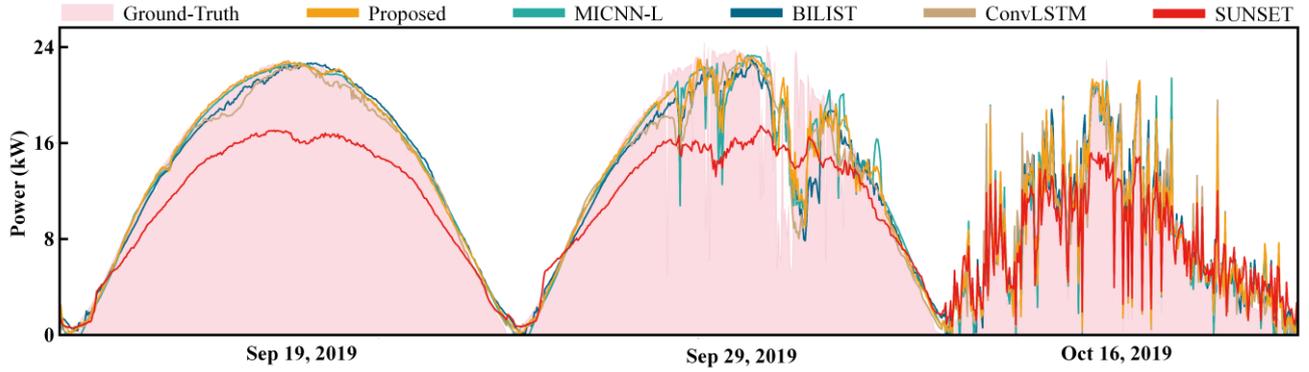

**Fig. 8 Forecast results for three selected days representing sunny, cloudy, and overcast weather conditions.**

### 3.4. Ablation study

This study performed systematic ablation experiments to evaluate the impact of different modalities on prediction performance, specifically the TAM, PAM, and VAM. The experiment was designed as follows: retaining only TAM; integrating TAM and PAM while excluding visual data; integrating TAM and VAM while excluding prompt information; and utilizing PAM and VAM. In addition, one extended setting was tested: the full model without soft prompt information.

Table 4 Ablation study for power forecasting results. For RMSE and MAE, lower values indicate better performance, while a higher $R^2$ indicates better performance. Red: the best, Blue: the second best.

| Methods | Horizon | RMSE(kW) | MAE(kW) | $R^2$(/) |
|---|---|---|---|---|
| Proposed | 20 | 1.603 | 0.698 | 0.948 |
|  | 40 | 1.766 | 0.857 | 0.937 |
|  | 60 | 1.925 | 0.947 | 0.925 |
| Proposed without soft prompt | 20 | 1.619 | 0.688 | 0.947 |
|  | 40 | 1.813 | 0.845 | 0.934 |
|  | 60 | 1.928 | 0.969 | 0.925 |
| TAM | 20 | 2.145 | 1.581 | 0.907 |
|  | 40 | 2.21 | 1.584 | 0.902 |
|  | 60 | 2.208 | 1.305 | 0.902 |
| TAM-PAM | 20 | 1.762 | 1.065 | 0.938 |
|  | 40 | 1.92 | 1.043 | 0.926 |
|  | 60 | 2.068 | 1.123 | 0.914 |
| TAM-VAM | 20 | 1.733 | 0.990 | 0.940 |
|  | 40 | 1.874 | 1.097 | 0.929 |
|  | 60 | 2.153 | 1.343 | 0.907 |
| PAM-VAM | 20 | 7.049 | 6.164 | 0 |
|  | 40 | 6.991 | 6.084 | 0.015 |
|  | 60 | 5.767 | 4.888 | 0.330 |

The results in the Table 4 show that the complete model (Proposed), which integrates temporal, textual, and visual modalities, consistently delivers the best performance in terms of RMSE, MAE, and $R^2$, particularly for the 40- and 60-minute forecasts. The variant that omits the soft prompt performs nearly as well, suggesting that while the soft prompt provides some minor improvements, the primary gains come from the full multimodal fusion. Using only TAM results in considerably higher errors and lower R² values, and although combining TAM with either textual or visual information improves the results compared to using TAM alone, these configurations still do not reach the performance level of the complete model. In contrast, the configuration that employs only PAM and VAM, excluding temporal information, shows the weakest performance, underscoring the essential role of time-based features. Overall, the findings highlight that temporal information is crucial for capturing the dynamic changes in PV power, and that the complementary contributions of textual and visual inputs further enhance prediction accuracy and robustness.

### 3.5. Transfer Study

In this study, to assess the transferability of the model in cross-dataset forecasting scenarios, the model was trained on Dataset A and then directly deployed on Dataset B without fine-tuning. This cross-domain evaluation enables us to examine whether the model can maintain high performance when applied to PV plants under different climate conditions and in varying locations, thereby retaining high levels of prediction accuracy and robustness. Moreover, we conducted tests on two distinct plants within Dataset B, each representing different climatic and geographic settings: Location 1 is Building 28 on the Main Campus of the University of Wollongong, with a peak PV output of approximately 30.84 kW, and Location 2 is the Sustainable Buildings Research Centre on the Innovation Campus of the same university, with a peak PV output of about 6.26 kW, located roughly 2 km from Location 1.

Table 5 Transfer study on power forecasting performance across different methods for Location 1. For RMSE and MAE, lower values indicate better performance, while a higher R² indicates better performance. Red: the best, Blue: the second best.

| Methods | Proposed | | | MICNN-L | | | BILIST | | | ConvLSTM | | | SUNSET | | |
|---|---|---|---|---|---|---|---|---|---|---|---|---|---|---|---|
| Horizon | 20 | 40 | 60 | 20 | 40 | 60 | 20 | 40 | 60 | 20 | 40 | 60 | 20 | 40 | 60 |
| RMSE($10^4$kW) | 0.251 | 0.294 | 0.349 | 0.270 | 0.396 | 0.351 | 0.366 | 0.537 | 0.458 | 0.459 | 0.320 | 0.426 | 1.51 | 1.52 | 1.42 |
| MAE($10^4$kW) | 0.173 | 0.189 | 0.235 | 0.167 | 0.345 | 0.259 | 0.271 | 0.417 | 0.337 | 0.374 | 0.221 | 0.270 | 1.39 | 1.39 | 1.26 |
| $R^2$(/) | 0.850 | 0.793 | 0.705 | 0.827 | 0.624 | 0.701 | 0.682 | 0.308 | 0.492 | 0.499 | 0.755 | 0.559 | 0 | 0 | 0 |

Table 6 Transfer study on power forecasting performance across different methods for Location 2. For RMSE and MAE, lower values indicate better performance, while a higher R² indicates better performance. Red: the best, Blue: the second best.

| Methods | Proposed | | | MICNN-L | | | BILIST | | | ConvLSTM | | | SUNSET | | |
|---|---|---|---|---|---|---|---|---|---|---|---|---|---|---|---|
| Horizon | 20 | 40 | 60 | 20 | 40 | 60 | 20 | 40 | 60 | 20 | 40 | 60 | 20 | 40 | 60 |
| RMSE($10^3$kW) | 0.588 | 0.681 | 0.826 | 0.672 | 0.721 | 0.902 | 0.744 | 0.956 | 0.951 | 0.798 | 0.790 | 0.971 | 2.63 | 2.35 | 2.06 |
| MAE($10^3$kW) | 0.404 | 0.475 | 0.564 | 0.486 | 0.474 | 0.663 | 0.534 | 0.761 | 0.699 | 0.545 | 0.583 | 0.665 | 2.33 | 2.02 | 1.77 |
| $R^2$(/) | 0.830 | 0.768 | 0.653 | 0.778 | 0.740 | 0.586 | 0.728 | 0.543 | 0.540 | 0.687 | 0.688 | 0.521 | 0 | 0 | 0 |

Table 5 and Table 6 present the transfer study results for power forecasting performance across different methods at Location 1 and Location 2, respectively. Across both locations, the proposed method demonstrates superior RMSE, MAE, and R² scores over most prediction horizons compared to the other approaches. Specifically, it achieves an average reduction of around 7% in RMSE, 9.5% in MAE relative to the next-best method, while also yielding a clear increase in $R^2$. On the 20, 40 and 60 minute horizons, it attains $R^2$ values of 0.850, 0.793 and 0.705 at Location 1 and 0.830, 0.768 and 0.653 at Location 2. Typically, $R^2$ above 0.80 indicates excellent predictive fit, values between 0.70 and 0.80 indicate strong fit and values above 0.65 remain acceptable under distributional shifts. These results demonstrate the robustness of the proposed method in adapting to distributional shifts between datasets, indicating that the features it learns under one set of conditions can effectively capture power fluctuation patterns in another. By contrast, other models show higher error rates and lower R² when deployed in cross-dataset scenarios, suggesting less adaptability. Overall, these results confirm both the transferability and generalization capability of the proposed method, underscoring its potential for practical PV power forecasting in diverse environments. As shown in Fig. 9, the proposed method produces forecast curves that more closely follow the ground-truth values than the baseline methods.

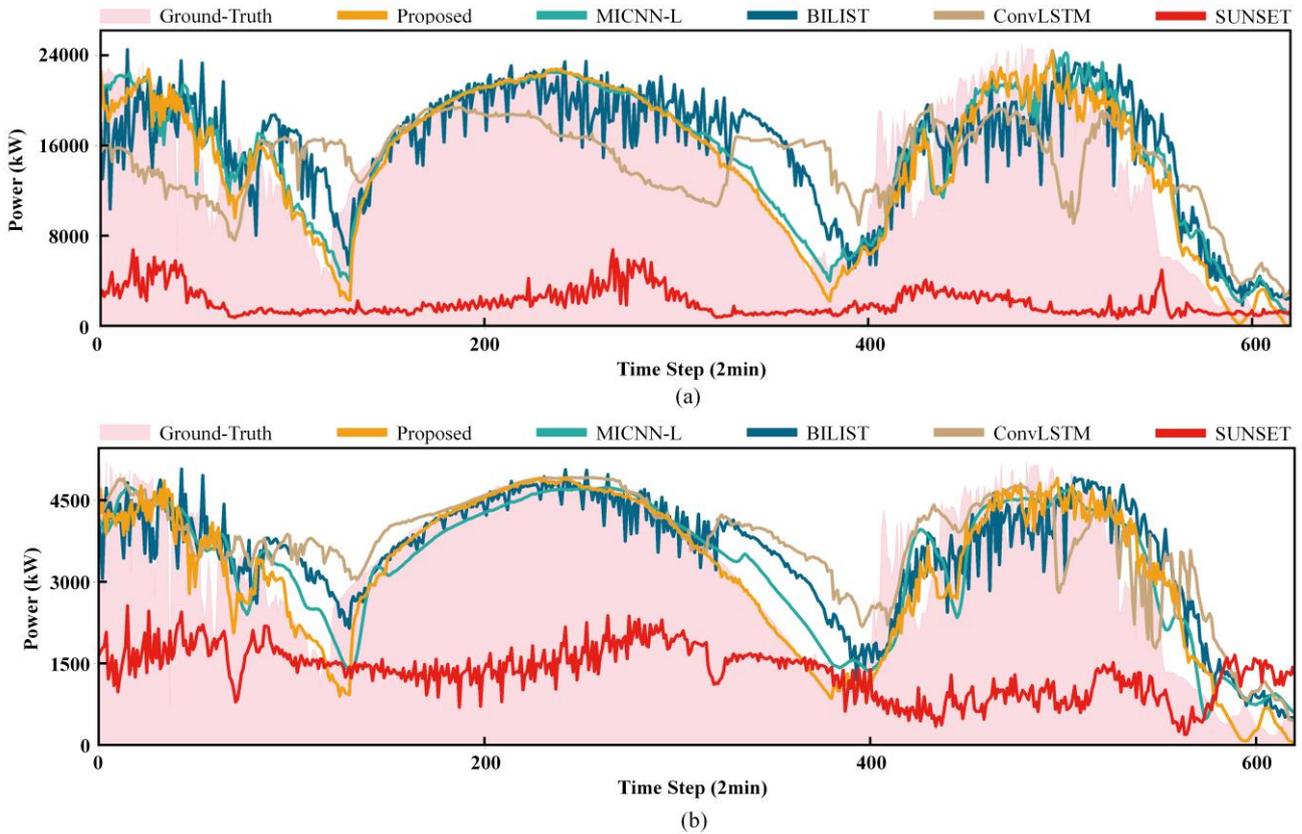

**Fig. 9 Cross-dataset power forecasting results for the proposed method and baseline models on 20-minutes horizon. (a) Location 1; (b) Location 2.**

## 4. Conclusion

In this study, a multimodal PV power forecasting framework, PV-VLM, was proposed to address the challenges of intra-hour photovoltaic power forecasting. By jointly incorporating temporal sequences, domain-specific prompt information, and sky image features, the model effectively captures the complex spatiotemporal dependencies and semantic patterns inherent in solar power generation. The design leverages a VLM for spatial feature extraction, a LLM for integrating prior knowledge through prompts, and a time series module for capturing local temporal dynamics.

Comprehensive experiments were conducted to evaluate the effectiveness of each component. Baseline results on Dataset A demonstrate that PV-VLM consistently outperforms existing baselines across multiple forecasting horizons, confirming the advantage of multimodal fusion. The ablation study further highlights the distinct role of each module: the temporal module captures evolving generation trends, the prompt module introduces contextual and expert-level priors, and the vision module contributes fine-grained sky condition representations. Additionally, the transfer study shows that the model trained on Dataset A retains strong forecasting capability when applied to the geographically distinct Dataset B, illustrating its robustness and generalization under varying climatic and spatial conditions. Comparative experiments indicate that the proposed model achieves an average reduction of approximately 5% in RMSE and nearly a 6% improvement in MAE across different forecasting horizons, while the transfer study further demonstrates average RMSE and MAE reductions of about 7% and 9.5%, respectively.

Future research will focus on extending the framework by incorporating additional environmental modalities, such as meteorological reports and satellite imagery, and enhancing the cross-modal fusion and alignment mechanisms to promote deeper semantic integration. These improvements are expected to contribute toward the development of more accurate and reliable solar forecasting systems, facilitating the integration of photovoltaic power into power grids under diverse real-world scenarios.

**Acknowledgments**

This work was supported by the National Natural Science Foundation of China under Grant No. 52077194.

**CRediT authorship contribution statement**

**Huapeng Lin:** Conceptualization, Methodology, Software, Formal analysis, Visualization, Validation, Investigation, Writing- Original Draft. **Miao Yu:** Conceptualization, Methodology, Writing- Review & Editing, Supervision.

**Declaration of competing interest**

The data that support the findings of this study are available from the authors on reasonable request.

**Competing interests**

The authors declare that they have no known competing financial interests or personal relationships that could have appeared to influence the work reported in this paper.

**References**

[1] Kabir E, Kumar P, Kumar S, Adelodun AA, Kim K-H. Solar energy: Potential and future prospects. Renewable and Sustainable Energy Reviews 2018;82:894–900. https://doi.org/10.1016/j.rser.2017.09.094.

[2] Wang F, Xu H, Xu T, Li K, Shafie-khah M, Catalão JoãoPS. The values of market-based demand response on improving power system reliability under extreme circumstances. Applied Energy 2017;193:220–31.

https://doi.org/10.1016/j.apenergy.2017.01.103.

[3] Hu P, Wang G, Tan Y-P. Recurrent Spatial Pyramid CNN for Optical Flow Estimation. IEEE Trans Multimedia 2018;20:2814–23. https://doi.org/10.1109/TMM.2018.2815784.

[4] Liu J, Zang H, Ding T, Cheng L, Wei Z, Sun G. Harvesting spatiotemporal correlation from sky image sequence to improve ultra-short-term solar irradiance forecasting. Renewable Energy 2023;209:619–31. https://doi.org/10.1016/j.renene.2023.03.122.

[5] Zhen Z, Pang S, Wang F, Li K, Li Z, Ren H, et al. Pattern Classification and PSO Optimal Weights Based Sky Images Cloud Motion Speed Calculation Method for Solar PV Power Forecasting. IEEE Trans on Ind Applicat 2019;55:3331–42. https://doi.org/10.1109/TIA.2019.2904927.

[6] Karout Y, Thil S, Eynard J, Guillot E, Grieu S. Hybrid intrahour DNI forecast model based on DNI measurements and sky-imaging data. Solar Energy 2023;249:541–58. https://doi.org/10.1016/j.solener.2022.11.032.

[7] Eşlik AH, Akarslan E, Hocaoğlu FO. Short-term solar radiation forecasting with a novel image processing-based deep learning approach. Renewable Energy 2022;200:1490–505. https://doi.org/10.1016/j.renene.2022.10.063.

[8] Xu S, Zhang R, Ma H, Ekanayake C, Cui Y. On vision transformer for ultra-short-term forecasting of photovoltaic generation using sky images. Solar Energy 2024;267:112203. https://doi.org/10.1016/j.solener.2023.112203.

[9] Paulescu M, Blaga R, Dughir C, Stefu N, Sabadus A, Calinoiu D, et al. Intra-hour PV power forecasting based on sky imagery. Energy 2023;279:128135. https://doi.org/10.1016/j.energy.2023.128135.

[10] Fu Y, Chai H, Zhen Z, Wang F, Xu X, Li K, et al. Sky Image Prediction Model Based on Convolutional Auto-Encoder for Minutely Solar PV Power Forecasting. IEEE Trans on Ind Applicat 2021;57:3272–81. https://doi.org/10.1109/TIA.2021.3072025.

[11] Nie Y, Zamzam AS, Brandt A. Resampling and data augmentation for short-term PV output prediction based on an imbalanced sky images dataset using convolutional neural networks. Solar Energy 2021;224:341–54. https://doi.org/10.1016/j.solener.2021.05.095.

[12] Wu S, Zhang W, Xu L, Jin S, Li X, Liu W, et al. CLIPSelf: Vision Transformer Distills Itself for Open-Vocabulary Dense Prediction. The Twelfth International Conference on Learning Representations, ICLR 2024, Vienna, Austria, May 7-11, 2024, OpenReview.net; 2024. https://doi.org/10.48550/arXiv.2310.01403.

[13] Shi C, Yang S. EdaDet: Open-Vocabulary Object Detection Using Early Dense Alignment. Proceedings of the IEEE/CVF International Conference on Computer Vision, 2023, p. 15724–34. https://doi.org/10.48550/arXiv.2309.01151.

[14] Jin S, Jiang X, Huang J, Lu L, Lu S. LLMs Meet VLMs: Boost Open Vocabulary Object Detection with Fine-grained Descriptors. The Twelfth International Conference on Learning Representations, ICLR 2024, Vienna, Austria, May 7-11, 2024, OpenReview.net; 2024. https://doi.org/10.48550/arXiv.2402.04630.

[15] Jia F, Wang K, Zheng Y, Cao D, Liu Y. GPT4MTS: Prompt-based Large Language Model for Multimodal Time-series Forecasting. Proceedings of the AAAI Conference on Artificial Intelligence 2024;38:23343–51. https://doi.org/10.1609/aaai.v38i21.30383.

[16] Cao D, Jia F, Arik SÖ, Pfister T, Zheng Y, Ye W, et al. TEMPO: Prompt-based Generative Pre-trained Transformer for Time Series Forecasting. The Twelfth International Conference on Learning Representations, ICLR 2024, Vienna, Austria, May 7-11, 2024, OpenReview.net; 2024. https://doi.org/10.48550/arXiv.2310.04948.

[17] Sun C, Li H, Li Y, Hong S. TEST: Text Prototype Aligned Embedding to Activate LLM's Ability for Time Series. The Twelfth International Conference on Learning Representations, ICLR 2024, Vienna, Austria, May 7-11, 2024, OpenReview.net; 2023. https://doi.org/10.48550/arXiv.2308.08241.

[18] Nie Y, Nguyen NH, Sinthong P, Kalagnanam J. A Time Series is Worth 64 Words: Long-term Forecasting with Transformers. The Eleventh International Conference on Learning Representations, ICLR 2023, Kigali, Rwanda, May 1-5, 2023, OpenReview.net; 2023. https://doi.org/10.48550/arXiv.2211.14730.

[19] Radford A, Kim JW, Hallacy C, Ramesh A, Goh G, Agarwal S, et al. Learning Transferable Visual Models From Natural Language Supervision. Proceedings of the 38th International Conference on Machine Learning, PMLR; 2021, p. 8748–63. https://doi.org/10.48550/arXiv.2103.00020.

[20] Li J, Li D, Savarese S, Hoi S. BLIP-2: Bootstrapping Language-Image Pre-training with Frozen Image Encoders and Large Language Models. Proceedings of the 40th International Conference on Machine Learning, PMLR; 2023, p. 19730–42. https://doi.org/10.48550/arXiv.2301.12597.

[21] Tschannen M, Gritsenko A, Wang X, Naeem MF, Alabdulmohsin I, Parthasarathy N, et al. SigLIP 2: Multilingual Vision-Language Encoders with Improved Semantic Understanding, Localization, and Dense Features 2025. https://doi.org/10.48550/arXiv.2502.14786.

[22] Vaswani A, Shazeer N, Parmar N, Uszkoreit J, Jones L, Gomez AN, et al. Attention is all you need. Proceedings of the 31st International Conference on Neural Information Processing Systems, Red Hook, NY, USA: Curran Associates Inc.; 2017, p. 6000–10. https://doi.org/10.48550/arXiv.1706.03762.

[23] Jin M, Wang S, Ma L, Chu Z, Zhang JY, Shi X, et al. Time-LLM: Time Series Forecasting by Reprogramming Large Language Models. The Twelfth International Conference on Learning Representations, ICLR 2024, Vienna, Austria, May 7-11, 2024, OpenReview.net; 2024.

[24] Liu X, Ji K, Fu Y, Tam WL, Du Z, Yang Z, et al. P-Tuning v2: Prompt Tuning Can Be Comparable to Fine-tuning Universally Across Scales and Tasks 2022. https://doi.org/10.48550/arXiv.2110.07602.

[25] Radford A, Wu J, Child R, Luan D, Amodei D, Sutskever I. Language Models are Unsupervised Multitask Learners. OpenAI 2019. https://cdn.openai.com/better-language-models/language_models_are_unsupervised_multitask_learners.pdf.

[26] Devlin J, Chang M-W, Lee K, Toutanova K. BERT: Pre-training of Deep Bidirectional Transformers for Language Understanding. In: Burstein J, Doran C, Solorio T, editors. Proceedings of the 2019 Conference of the North American Chapter of the Association for Computational Linguistics: Human Language Technologies, NAACL-HLT 2019, Minneapolis, MN, USA, June 2-7, 2019, Volume 1 (Long and Short Papers), Association for Computational Linguistics; 2019, p. 4171–86. https://doi.org/10.18653/V1/N19-1423.

[27] Qwen, Yang A, Yang B, Zhang B, Hui B, Zheng B, et al. Qwen2.5 Technical Report 2025. https://doi.org/10.48550/arXiv.2412.15115.

[28] He K, Zhang X, Ren S, Sun J. Deep Residual Learning for Image Recognition. Proceedings of the IEEE Conference on Computer Vision and Pattern Recognition, 2016, p. 770–8. https://doi.org/10.48550/arXiv.1512.03385.

[29] Nie Y, Li X, Scott A, Sun Y, Venugopal V, Brandt A. SKIPP'D: A SKy Images and Photovoltaic Power Generation Dataset for short-term solar forecasting. Solar Energy 2023;255:171–9. https://doi.org/10.1016/j.solener.2023.03.043.

[30] Dissawa L, Robinson D, Agalgaonkar A, Godaliyadda R, Ekanayake P, Perera S, et al. Sky Images and PV Power Measurements for Irradiance Forecasting 2021. https://doi.org/10.17632/cb8t8np9z3.2.

[31] Ruiz-Arias JA, Gueymard CA. CAELUS: Classification of sky conditions from 1-min time series of global solar irradiance using variability indices and dynamic thresholds. Solar Energy 2023;263:111895. https://doi.org/10.1016/j.solener.2023.111895.

[32] Ajith M, Martínez-Ramón M. Deep learning based solar radiation micro forecast by fusion of infrared cloud images and radiation data. Applied Energy 2021;294:117014. https://doi.org/10.1016/j.apenergy.2021.117014.

[33] Zhang R, Ma H, Saha TK, Zhou X. Photovoltaic Nowcasting With Bi-Level Spatio-Temporal Analysis Incorporating Sky Images. IEEE Trans Sustain Energy 2021;12:1766–76. https://doi.org/10.1109/TSTE.2021.3064326.

[34] Nie Y, Paletta Q, Scott A, Pomares LM, Arbod G, Sgouridis S, et al. Sky image-based solar forecasting using deep learning with heterogeneous multi-location data: Dataset fusion *versus* transfer learning. Applied Energy 2024;369:123467. https://doi.org/10.1016/j.apenergy.2024.123467.

[35] Sun Y, Venugopal V, Brandt AR. Short-term solar power forecast with deep learning: Exploring optimal input and output configuration. Solar Energy 2019;188:730–41. https://doi.org/10.1016/j.solener.2019.06.041.

[36] Zang H, Chen D, Liu J, Cheng L, Sun G, Wei Z. Improving ultra-short-term photovoltaic power forecasting using a novel sky-image-based framework considering spatial-temporal feature interaction. Energy 2024;293:130538. https://doi.org/10.1016/j.energy.2024.130538.